# Security-aware selection of Web Services for Reliable Composition

Shahedeh A.khani
Centre for Software Reliability,
City University,
London,
United Kingdom
abdolhossein.shahedeh.1@city.ac.uk

Cristina Gacek
Centre for Software Reliability,
City University,
London,
United Kingdom
cristina.gacek.1@city.ac.uk

Peter Popov
Centre for Software Reliability,
City University,
London,
United Kingdom
p.t.popov@city.ac.uk

*Abstract*— **Dependability is an important characteristic that a trustworthy computer system should have. It is a measure of Availability, Reliability, Maintainability, Safety and Security. The focus of our research is on security of web services. Web services enable the composition of independent services with complementary functionalities to produce value-added services, which allows organizations to implement their core business only and outsource other service components over the Internet, either pre-selected or on-the-fly. The selected third party web services may have security vulnerabilities. Vulnerable web services are of limited practical use. We propose to use an intrusion-tolerant composite web service for each functionality that should be fulfilled by a third party web service. The third party services employed in this approach should be selected based on their security vulnerabilities in addition to their performance. The security vulnerabilities of the third party services are assessed using a penetration testing tool. In this paper we present our preliminary research work.**

## I. INTRODUCTION

Web Services (WSs) are used to implement Service Oriented Architectures (SOAs). Each service consists of an implementation that is on a network-accessible platform and an interface. The communication with WSs is supported by Simple Object Access Protocol (SOAP) [1], which is a standard protocol for packaging messages before transmitting them through standard Internet technologies. Web Service Description Language (WSDL) [2] is used to describe the interface of a WS, describing the concrete structure of the SOAP messages (description of the operations and their input and output parameters), the service's protocol binding and network location for the WS's implementation (e.g. the URL). SOAP messages and WSDL documents are based on eXtensible Markup Language (XML), which has simplified the interoperability of the various technologies employed in WS development.

WSs enable the composition of independent services with complementary functionalities to produce value-added services, which allows organizations to implement their core business only and outsource other service components over the Internet, either pre-selected or on-the-fly. However, WSs are open to a large population of users therefore, maintaining their security is an important task. Security attacks on WSs may cause unavailability and/or loss of confidentiality and integrity as well as significant monetary penalties. Now the question is what if the selected third party WSs have security vulnerabilities? Vulnerable WSs are of limited practical use. Therefore, security should also be considered in selection and employment of suitable third party WSs.

Our objective is to improve the dependability of composite WSs in which third party WSs are used. Our focus is on the security of WSs. Web services are at risk of security vulnerabilities related to their specific implementation technologies (e.g. XML) as well as those of their underlying platforms (e.g. operating systems and web-services frameworks) and the WS applications themselves (e.g. being vulnerable to SQL injection attacks). Security vulnerabilities related to WSs' implementation technologies are central to the work described in this paper. We propose to use an intrusion-tolerant composite WS (using fault tolerant techniques: N-version programming, diversity) for each functionality that should be fulfilled by a third party WS. In our approach, penetration testing is used to identify security vulnerabilities of available functionally-equivalent candidate third party WSs. Suitable third party WSs will then be selected based on their security vulnerabilities and their performance according to the client's (being the owner of the system) requirements. The selected services will be invoked using an intrusion-tolerant approach as a countermeasure against the security attacks exploiting the vulnerabilities that are not covered/identified by penetration testing.

This paper makes the following contributions:

- It explains our proposed approach in details.
- It exemplifies our approach using a case study.

The remainder of this paper is organised as follows. Related work is discussed in Section II. Section III briefly introduces WSs' specific security vulnerabilities. Penetration testing and the tool we are using are discussed Section IV. Sections V and VI present our proposed approach and the case study respectively. Section VII draws the conclusions and outlines future work.

## II. RELATED WORK

Various intrusion detection and prevention methods and approaches have been proposed to secure WSs. However, in addition to the adoption of the proposed methods and approaches, WSs should also be able to tolerate attacks and continue to provide an acceptable level of service even after intruders have broken in. Intrusion-tolerant systems can be developed using fault-tolerance concepts and approaches.

Redundancy is believed to be a valid defence against physical faults. Running multiple replicas of the system and switching to the functioning one when a failure occurs, is an example of using redundancy to overcome hardware faults [3]–[5]. Redundancy can also be applied to the code, data, and environment of a software system to overcome its nonphysical faults [6]. Design diversity is a recognised defence against design faults. Littlewood, Popov and Strigini [7] have surveyed the benefits of design diversity. Carzaniga, Gorla and Pezzè [8] describe the redundancy as a system's capability of executing the same functionality in several execution environments or in various ways (e.g. using different execution paths). Littlewood and Strigini [9] have argued the validity of using redundancy and diversity for security.

Majorczyk et al. [10] have proposed Intrusion Detection Systems (IDS) based on redundancy and diversification of Components-Off-The-Shelf (COTS) and have applied it to web servers. Valdes et al. [11] have proposed an intrusion-tolerant web server architecture based on redundant COTS servers running on diverse operating systems and platforms. Gorbenko et al. [12] have proposed a generic intrusion-avoidance architecture to be used for deploying WSs in the cloud. This architecture employs software diversity at various system levels and dynamically reconfigures the cloud deployment environment. All above approaches use redundancy and diversity techniques to detect and tolerate the intrusions. However, none of them addresses the attacks exploiting the vulnerabilities caused by the XML standards, which are independent of the type of operating systems or application implementation. Massimo Ficco and Massimiliano Rak [13] have proposed an intrusion tolerance approach for DoS attacks to WSs. It focuses on the detection of attack symptoms, as well as the diagnosis of intrusion effects in order to take appropriate action only when the attack succeeds. This work is more related to our approach. However, it focuses on a specific group of XML DoS attacks, called Deeply-Nested XML.

## III. WEB SERVICES' SPECIFIC SECURITY VULNERABILITIES

As stated previously, the communication between WSs is supported by XML-based protocols. This makes WSs vulnerable to XML attacks (vulnerabilities related to their specific implementation technologies). Examples of recently reported security attacks exploiting such vulnerabilities are the attacks on Amazon EC2 SOAP, Eucalyptus cloud WS interfaces [14], [15], different SAML-based frameworks [16] and ciphertext decryption exploitation [17]. Jensen et al. [18], [19] present a list of top WSs' specific security vulnerabilities (related to the implementation technologies). To identify these security vulnerabilities, they have performed exemplary attacks on widespread WS implementations. According to the study, some of these vulnerabilities are due to implementation weaknesses but majority of them are due to protocol flaws. In this section we briefly introduce a number of these security vulnerabilities.

### A. Attack Obfuscation

WS-Security [20] is a very flexible security standard that allows signing and encrypting only parts of the message, which contains sensitive data. A disadvantage of using this standard is that the encrypted content may not be inspected without prior decryption. Such encryption can be used by attackers to conceal malicious code. Therefore, if the encrypted part of the message contains an intended attack (e.g. Denial of Service attack.), it will be very difficult to detect.

### B. XML Injection

An XML Injection attacker tries to modify the structure of a XML document (e.g. SOAP message) by adding some contents containing XML tags.

### C. SOAPAction Spoofing

A SOAP message package consists of a transport protocol header and an envelope. The SOAP envelope consists of a header and a body. The first child element of the body contains the operation addressed by the SOAP request [18]. If the HTTP transport protocol is used, an additional operation identifier element called SOAPAction can be added to the header [18]. This enables the receiving WS to understand what operation the SOAP body contains, prior to XML parsing [21]. However, it is often used as the only qualifier for the requested operation [22].

A WS will be vulnerable to a SOAPAction spoofing attack if the requested operation is identified solely based on the SOAPAction value or first child element of the SOAP body [22]. A successful SOAPAction Spoofing attack will result in unauthorised execution of operations offered by the WS.

### D. Denial of Service (DoS) Attacks

Early steps in processing a request SOAP message include parsing and transforming the contents of the message to be usable by the WS's backend applications. Therefore, the XML parser is an essential part of a WS's application logic. Simple API for XML (SAX) [23] and Document Object Model (DOM) [24] are two typical XML parsers.

DOM parsers read the whole XML stream into the memory then create hierarchical objects for each node (an element, an attribute etc.), referenced by the application logic. An attacker can plot a DoS attack on a DOM-based WS by inputting a large XML file [25]. Such attacks (e.g. Oversize payload and Coercive parsing [19]) affect the availability of the WS by exhausting its resources and preventing legitimate users from accessing the service [26]. DOM parsers can also be subject to other types of attacks such as XML injection [25].

On the other hand, SAX parsers perform XML parsing at the start or end of a node without loading the whole XML stream into memory (they load a maximum of two elements into the memory at a time) [25]. Whenever the parser reaches a node, it triggers an event, and the program's event handler starts processing the data. SAX-based WSs are vulnerable to XML injection attacks [25]. In XML injection attacks the attacker targets the integrity of the XML stream (e.g. SOAP message) by overwriting static portions of it [25]. For example, the attacker modifies the message by adding contents containing XML tags [27].

DoS attacks are one of the most popular attacks, which can be performed through a variety of techniques. This type of attacks exploit the vulnerabilities in XML-based documents (e.g. SOAP messages) targeting the parsing mechanisms and other resources, affecting the availability of the WS. A large number of these attacks, targeting well-known companies such as VISA and PayPal suggests that they can be a serious threat to today's IT infrastructure [28]. Jensen *et al.* [18], [19] have presented a number of DoS attacks.

### E. Hash Collision (HashDoS) Attack

Hash tables can be employed within a SOAP message to store values and their references (e.g. attributes and their corresponding namespace). Ideally each key should represent a unique value. If different keys represent the same value, a collision will happen, which results in resource intensive computation. An attacker can exploit a weak hash function to perform a DoS attack [29], [30].

## IV. PENETRATION TESTING

Penetration testing or static code analysis approaches can be employed to assess the security of a WS [31]. Penetration testing is an attempt to break into a system not in order to exploit it, but rather to identify its weaknesses [32]. The resistance of the system against penetration testing is a good indicator of its security [27]. In our approach, security vulnerabilities of a third party service play an important role in its selection. Security vulnerabilities related to WSs' implementation technologies are central to our work. Hence, a penetration testing tool called WS-Attacker [27] is chosen for testing candidate third party WSs to identify their security vulnerabilities. The reasons for WS-Attacker selection are: (1) it enables testing for XML-specific security vulnerabilities explained in Section III (2) it performs the attacks automatically and (3) it is an open source penetration testing tool.

WS-Attacker consists of a framework and plugins architecture. Its framework is based on soapUI [32] and sets up an environment for attacking WSs. In WS-Attacker, the attacks are implemented as plugins. Each plugin is an implementation of a model of an adversary performing one type of attack and allows the user to set various parameters, such as number of parallel attack threads, number of requests per thread, milliseconds between every test-probe requests, and milliseconds between every attack requests. WS-Attacker is extendable and provides a plugin interface enabling new attack plugins to be added to the tool [27]. A number of attack plugins have been developed for WS-Attacker by its developers and other researchers [27], [33], [28]. In implementing these plugins, the developers assume that the tester does not have a direct access to the system under attack, and can only examine its vulnerability to an attack by sending payloads to its server then evaluating its response (or its response time in the case of performing DoS attacks). The result (true or false) of performing an attack indicate whether it has been successful [27].

## V. PROPOSED APPROACH

As stated previously, WSs are at risk of security vulnerabilities related to their specific implementation technologies (e.g. XML) as well as those of their underlying platforms (e.g. operating systems and web-services frameworks) and the WSs applications themselves (e.g. being vulnerable to SQL injection attacks). Systems could be developed using services offered by different vendors. To explain our approach, we are assuming that we have control over a system that is under development, which should employ third party WSs. The candidate third party WS(s) may have security vulnerabilities. Their security vulnerabilities may be exploited by messages from our system's client, the clients of other systems that are also employing them or their direct clients. The first scenario will definitely affect our system. The last two scenarios could affect our system as a result of security attacks, such as DoS, which may make the third party service unavailable just for few minutes or completely with the need to reboot.

We propose to use an intrusion-tolerant composite WS (using fault tolerant techniques: N-version programming, diversity) for each functionality that should be fulfilled by a third party WS. Our proposed approach includes the following steps:

Step 1. Collect as many WSs as possible, offered by various vendors providing the same functionality as required by the system.

Step 2. Log the failure rate of each service stated in its Service Level Agreement (SLA). Initially the failure rates stated by the provider of the third party service will be considered but this log will be updated every time the service is invoked.

Step 3. Identify their security vulnerabilities using a penetration testing tool (as shown in Fig. 1).

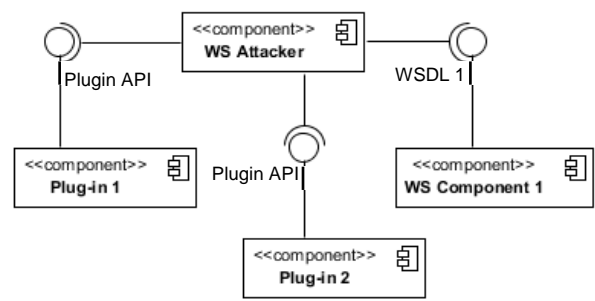

**Figure 1: Penetration Testing WSs.**

Step 4. Log the vulnerabilities of each service.

Step 5. Create a composite WS out of all available candidate WSs for the functionality that is to be outsourced.

- Step 6. For each request received from a client of the system, scan the message to identify which security vulnerabilities it could exploit if it contains malicious contents.
- Step 7. Select all third party services without vulnerabilities, which could be exploited if client's request contains malicious contents. The system understands which services should not be selected from the log generated in Step 4. For example a message with SOAPAction elements will exploit the security vulnerability of a third party WS to SOAPAction Spoofing attack if it contains malicious contents. Therefore, no services vulnerable to this type of attack should be selected if the client's request contains SOAPAction elements. The number of the selected WSs at this stage should be greater than or equal to $3f+1$ ($f$ is the number of possible faulty replicas). Castro and Liskov [34], have justified the optimality of $3f+1$ replicas. When the number of WSs without tested security vulnerabilities that could be exploited if client's request contains malicious contents is less than $3f+1$, select the remaining ones from WSs with security vulnerabilities that are of lowest priority to system owner and with lowest failure rates (using the log generated in Step 2).
- Step 8. If the number of services selected in Step 7 is greater than $3f+1$, select the services with lowest failure rates (using the log generated in Step 2).
- Step 9. Invoke WSs selected in Step 8 concurrently.
- Step 10. Re-execute the WSs if all of them fail (Active+Time replication strategy [35]).
- Step 11. Terminate the execution of WSs as soon as $2f+1$ responses are returned.
- Step 12. Applying majority voting to identify the response that should be returned to the client of the system.

## VI. CASE STUDY

To exemplify our approach we are using a stock purchase service. These experiments are run on Intel® Core™ i5-3320M CPU @ 2.60GHz system with 7.88GB usable RAM and 64-bit Operating System. In these experiments only security vulnerability to Coercive parsing attack is considered and the Steps of our proposed approach are taken as follows:

1. We have developed four WSs (two using Apache Axis2 and two using ASP.NET WS frameworks) and selected a third party ASP.NET WS, which provides similar functionality to those we have developed.
2. No information related to failure rate of these WSs was available. Hence, Step 2 of our approach is omitted in this experiment but it will be considered in our future work.
3. Each WS was tested individually for security vulnerability to Coercive Parsing attack by submitting the location of its WSDL file to WS-Attacker then performing the attack with settings (default settings) shown in Table 1.
4. The security vulnerability of these WSs to Coercive Parsing attack was identified (see Table 2). As it is explained in section IV, the developers of WS-Attacker's DoS attack plugins have assumed that the tester does not have direct access to the system under attack, and can only examine its vulnerability to the above DoS attacks by sending payloads to its server then evaluating its response time. They have defined the response time as the time when the last byte of the request is sent to the server until the first byte of the response is received from the server [33]. In designing these attack plugins, all major errors, such as increase in response time caused by variable message sizes or network loads, are eliminated [33]. These attack plugins calculate the median of the response times of the last 10 tampered requests and the median of the last 10 untampered requests. They then work out the ratio of the median response time of the tampered requests to the median response time of the untampered requests. Any ratio notably higher or lower than 1, will be interpreted as a successful attack. Refer to the results presented in Table 2, 100% indicates that the WS has vulnerability to Coercive Parsing attack and 1% shows that it has not this vulnerability.
5. A Business Process Execution Language (BPEL [36]) composite WS (shown in Fig. 2) was created using all all available WSs but only the three ASP.NET services and one of the Axis2 services (four services highlighted in Table 2) were invoked, addressing Steps 5-8 of proposed approach. Steps 6-8 should be performed automatically, which will be addressed in our future work.
6. To address the Steps 9 and 11 of our approach, all four services were invoked concurrently and a variable was dedicated to each of the concurrent processes. Upon receiving a response from each of the services, the corresponding variable would be set to a pre-defined value to indicate that a response is returned. A throw and catch exception handling was employed to terminate WSs execution upon receiving the first three responses (as soon as the dedicated variables to three out of four services are set to the pre-defined value).

The composite service was then tested for security vulnerability to Coercive Parsing attack, using WS-Attacker with the same settings that were used to test individual services. As the results from this experiment illustrate (last column of Table 2), the composite service is no longer vulnerable to Coercive Parsing DoS attack.

The purpose of this experiment was to exemplify our approach as well as testing its effectiveness as a defence against DoS attacks. However, our future work will (a) take into account the failure rate of the WSs (b) scan the client's request (c) performs Step 6-12 automatically (d) employ majority voting.

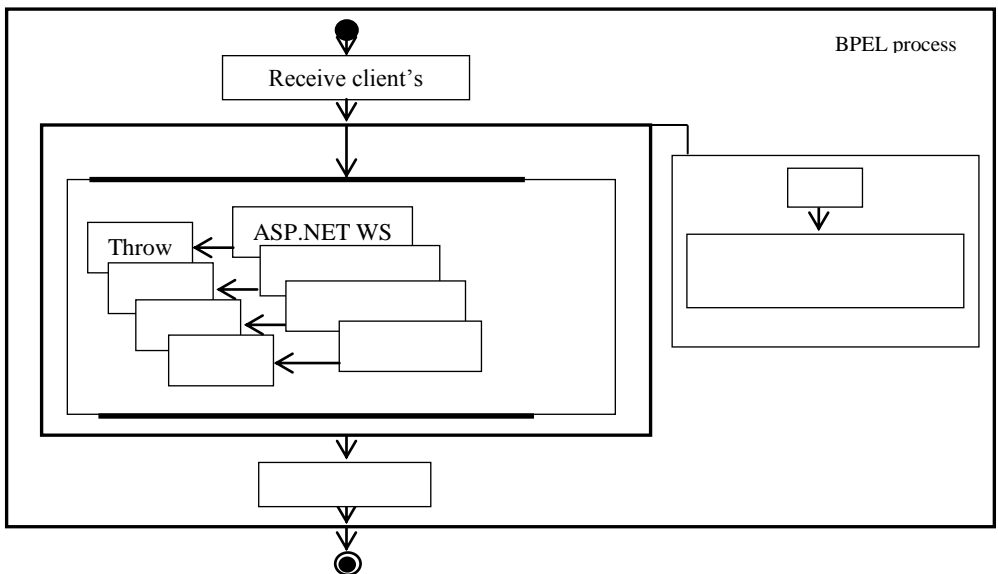

**Figure 2: BPEL Diagram of the Composite Service.**

**Table 1: Coerceive Parting Attack Settings Used for Experiments**

| Security Vulnerability Type | Common WS-Attacker settings | Specific WS-Attacker settings |
|---|---|---|
| Coercive Parsing | 2 parallel attack threads,<br>4 requests per thread,<br>500 milliseconds between every testprobe request,<br>750 milliseconds between every attack request,<br>4 seconds server recovery time,<br>5 seconds stop after the last tampered request. | 75,000 nested elements |

**Table 2: Preliminary Experimental Results**

| Security Vulnerability Type \ Web Service Framework | Axis2 web service | Axis2 web service | ASP.Net web service | ASP.Net web service | ASP.Net web service | Composite service |
|---|---|---|---|---|---|---|
| Coercive Parsing | 100% | 100% | 1% | 1% | 1% | 1% |

## VII. CONCLUSIONS

In this paper we proposed to use an intrusion-tolerant composite WS (using fault-tolerant techniques: N-version programming, diversity) for each functionality that should be fulfilled by a third party WS to improve the security of WSs. In our approach, penetration testing is used to identify security vulnerabilities of available functionally-equivalent candidate third party WSs. Suitable third party WSs will then be selected based on their security vulnerabilities and their performance according to the client's (being the owner of the system) requirements. The selected services will be invoked using an intrusion-tolerant approach as a countermeasure against the security attacks exploiting the vulnerabilities that are not covered/identified by penetration testing. We have presented our preliminary experimental results indicating that an intrusion-tolerant composite service may reduce the security vulnerabilities of WS.

Despite using an intrusion-tolerant composite service, we depend on a single orchestration engine, in these experiments BPEL, to manipulate them. Regardless of which orchestration engine we choose, it will always become a possible single point of failure. Thus, if the orchestration engine is vulnerable itself, then it may compromise any composite service that it handles. We note that in our experiments we did not observe this phenomenon. Standard solutions to address this concern would require applying-fault tolerance to the orchestration engines themselves, a problem which we will address in our future work.

We recognise that using the proposed approach introduces an additional delay in processing the requests sent to the BPEL orchestration in comparison to the clients sending the request directly to the component WS. Indeed, with the BPEL orchestration, the client request travels first to the BPEL engine and then it is forwarded to each of the component WSs. The resulting delay may increase significantly and will depend on the quality and speed of the connections.

The use of a composite service based on the best functionally equivalent set of diverse WSs is certainly beneficial for many security concerns (e.g., unavailability given DoS attacks). Nonetheless, if one or more of the underlying WSs has succumbed to an intruder, giving them access to incoming requests, the use of the composite service may actually make us more vulnerable to loss of confidentiality, as the requests will be sent to all of the underlying services, and not just one of them. Future work should explore how to reduce and/or mitigate this risk. In particular the adjudicator (performing majority voting) will play an important role here. We intend to study its impact systematically.

Currently we are investigating the effectiveness of our approach using WSs developed based on various web services frameworks. In the near future, we will focus on the implementation of a complete composition framework supporting security-aware service selection, composition and adaptation.